# Sub-millihertz magnetic spectroscopy with a nanoscale quantum sensor

**Authors:** Simon Schmitt[1], Tuvia Gefen[2], Felix M. Stürner[1], Thomas Unden[1], Gerhard Wolff[1], Christoph Müller[1], Jochen Scheuer[1,3], Boris Naydenov[1,3], Matthew Markham[4], Sebastien Pezzagna[5], Jan Meijer[5], Ilai Schwarz[3,6], Martin Plenio[3,6], Alex Retzker[2], Liam P. McGuinness[*,1], Fedor Jelezko[1,3]

**Affiliations:** [1] Institute of Quantum Optics, Ulm University, 89081 Ulm, Germany

[2] Racah Institute of Physics, Hebrew University of Jerusalem, 91904 Jerusalem, Israel

[3] Center for Integrated Quantum Science and Technology (IQST), Ulm University, 89081 Ulm, Germany

[4] Element Six, Harwell Campus, Fermi Avenue, Didcot, OX11 0QR, United Kingdom

[5] Institut für Experimentelle Physik II, Universität Leipzig, 04103 Leipzig, Germany

[6] Institute for Theoretical Physics, Ulm University, 89081 Ulm, Germany

**Email:** liam.mcguinness@uni-ulm.de

**Abstract:** Precise timekeeping is critical to metrology, forming the basis by which standards of time, length and fundamental constants are determined. Stable clocks are particularly valuable in spectroscopy as they define the ultimate frequency precision that can be reached. In quantum metrology, where the phase of a qubit is used to detect external fields, the clock stability is defined by the qubit coherence time, which determines the spectral linewidth and frequency precision. Here we demonstrate a quantum sensing protocol where the spectral precision goes beyond the sensor coherence time and is limited by the stability of a classical clock. Using this technique, we observe a precision in frequency estimation scaling in time $T$, as $T^{-3/2}$ for classical oscillating fields. The narrow linewidth magnetometer based on single spins in diamond is used to sense nanoscale magnetic fields with an intrinsic frequency resolution of 607μHz, 8 orders of magnitude narrower than the qubit coherence time.

**One Sentence Summary:** Combining a single spin sensor in diamond with a classical clock we demonstrate sub-millihertz nanoscale magnetic spectroscopy beyond the limits imposed by decoherence.

**Main Text:** Quantum sensing and metrology exploit physical laws governing individual quantum systems or correlations between systems to perform detection at the limits of sensitivity, resolution and non-invasiveness. In general, the techniques proceed by subjecting a quantum coherent probe to a time dependent Hamiltonian $H(t)$ which alters the evolution of the probe state over a duration limited by the probe coherence time $T_2$. A measurement is then performed to gain information about $H(t)$, e.g. its amplitude (Fig. 1A). As a result the probe coherence time sets a limit on sensor performance, since this is the maximum time over which coherent signal accumulation may be obtained. Subsequent measurements up to a total time $T$, may also be performed to improve the parameter estimation. For a single measurement with a sensor with long coherence time, a 'Heisenberg' $1/T$ scaling in the precision of amplitude estimation can be obtained (*1*). However, most algorithms achieve a precision which scales as $\frac{1}{\sqrt{T_2 T}}$, resulting in a sensitivity improving as $\sqrt{\text{Hz}}$ (*2-6*).

In addition to determining amplitude components, quantum spectroscopy aims at reconstructing the frequency spectrum of a given Hamiltonian as faithfully as possible. For example by minimising the error $\Delta\nu$, between an estimated frequency $\tilde{\nu}$ and a given frequency $\nu$ present in $H(t)$; which we refer to here as measurement precision. The current gold-standard techniques focus on adjusting the system Hamiltonian during the interaction period in order to isolate individual frequency components. Referred to as dynamical decoupling, such techniques involve coherently pulsing or driving the sensing qubit, resulting in the generation of a spectral filter with linewidth given by the inverse of the interaction time (Fig. 1B). The sensor linewidth is a critical figure of merit for spectral analysis as this defines not only the precision with which an individual frequency component can be estimated, but also the ability to resolve several frequencies. As for amplitude estimation, square root precision scaling is observed for frequency estimation, albeit with significant overhead due to the need to sequentially sample different frequencies when obtaining a complete spectrum (Fig. 1B,C).

Recently, the first approaches to reach a frequency resolution beyond the phase coherence time of the sensor qubit were performed. The protocols operated by transferring the quantum phase of the sensor to state population which survives up to longer $T_1$ relaxation times (*7, 8*), or using a hybrid quantum system where an additional long-lived qubit acts as a more stable clock (*9-13*). These techniques effectively replace the coherence time of the sensor qubit $T_2$, with the lifetime of a memory qubit $T_M$. Signal accumulation still requires a coherent sensor, and as a result the overall sensitivity is reduced when the memory time greatly exceeds the sensor period, since less time is spent acquiring a signal. Consistent with previous methods, the precision of frequency estimation scales as $1/\sqrt{T}$, but notably the linewidth can be greatly reduced due to the use of memory qubits with longer lifetimes.

Here we introduce a quantum spectroscopy technique where the spectral resolution is independent of the sensor coherence time and an enhanced precision scaling of $T^{-3/2}$ is obtained. The technique is analogous to classical heterodyne detection in that an unknown signal is 'mixed' with a local oscillator, however in our protocol, the non-linear mixing element is a quantum coherent probe. This quantum heterodyne (Qdyne) detection scheme retains critical elements of both conventional heterodyne detection and quantum sensing to form a hybrid quantum-classical device that outperforms each technique in isolation. In particular, the frequency resolution of Qdyne is defined by the local oscillator stability, enabling the measurement linewidth to be narrowed beyond the $T_2$ limit of the quantum sensor. Unlike conventional heterodyne detection however, the sensitivity is independent of the local oscillator intensity, and is derived purely from the quantum sensor.

These properties can be seen by analysing the interaction of a spin qubit with an oscillating magnetic field of the form $H(t) = k\sigma_z \sin(2\pi\nu t + \Phi)$, where $k$ is the interaction strength, $\sigma_z$ is the Pauli spin-z operator, and $\Phi$ is an arbitrary phase. Accumulation of a relative phase between two of the sensor's spin levels, denoted here as $|0\rangle, |1\rangle$, after being brought into a coherent superposition state $1/\sqrt{2}\,(|0\rangle + |1\rangle)$ can be tailored to depend not only on the amplitude and frequency of the detected field, but also on the signal phase with respect to the start of the measurement (Fig. 2). Specifically, for an XY8 sequence when π-pulses are applied with time separation close to $1/2\nu$, the sensor acquires a net phase $\varphi = \frac{2kT_s}{\pi}\cos(\Phi)$, where $T_s$ is the interaction time (see supplementary information). Usually no emphasis is placed on the signal phase since the detection is repeated many times and thereby averaged over all realisations. The idea we present here is to show that under the conditions that the signal remains coherent across multiple measurements, and the output state of the sensor depends on the signal phase $\Phi$; then control over the timing of subsequent experiments with a stable oscillator can be used to imprint changes in the signal phase onto the sensor. As a result, correlations across measurements for times exceeding the sensor coherence time can be observed. The sensor phase $\varphi_n$ at each measurement time $T_n = nT_L$, is then $\varphi_n = \frac{2kT_s}{\pi}\cos(2\pi\delta[T_n - T_L] + \Phi)$, where $T_L$ is the sum of the interaction time and the readout time, and $\delta = \nu - \nu_{LO}$, where $\nu_{LO}$ is a function of $T_L$ (see supplementary information). With knowledge of $T_L$, a measurement of the beat frequency $\delta$, allows $\nu$ to be determined with an accuracy determined by the local oscillator stability and up to a sign ambiguity. The amplitude of the beat note, and thus the sensitivity, is proportional to the sensor phase accumulation which is set by $T_s \leq T_2$. Here we use a photon correlation technique which allows frequency components fluctuating at timescales of several megahertz to be sampled even though high fidelity readout of our sensor requires several thousand measurement repetitions taking ~ 1 ms (see supplementary information).

We demonstrate Qdyne with single nitrogen vacancy (NV) centers positioned a few nanometres below the diamond surface. Such NV centers are nanoscale magnetic field detectors (*14, 15*) capable of detecting the magnetic field from small spin ensembles (*16-18*) down to the level of individual molecules (*19*) and nuclei (*20, 21*). Whilst the magnetic sensitivity of the NV center is sufficient for single molecule magnetic spectroscopy to be performed, the spectral resolution is limited to $\geq 1$ kHz due to the ~1 ms spin coherence time (*22*). For nanoscale metrology, decoupling spectral resolution from the quantum coherence of such sensors is especially relevant since it has been shown that near surface NV sensors have shorter coherence times and thus poorer spectral resolution (*23-25*). A loss of spectral resolution also occurs for high sensitivity ensemble magnetometers as the NV density is increased (*26*).

A sketch of the experimental setup is shown in Fig. 3A,B. A confocal microscope with 532nm excitation allows single, shallow NV centers to be optically detected and initialised into the |0⟩ spin state. Microwaves delivered by a 20µm copper wire provide full NV spin manipulation. Collection of

fluorescence photons with an avalanche photodiode allows readout of NV spin state with a contrast of ~30%. The arrival time of individual photons is recorded with nanosecond resolution by a time-tagged single photon counting card and stored to computer memory with a time-stamp synchronised to a local oscillator. The same local oscillator determines the timing $T_n$, of subsequent quantum measurements, providing a timing accuracy given by the clock stability (Fig 3A, supplementary information).

To illustrate the frequency resolution of Qdyne we applied an 880 nT magnetic field oscillating near 1 MHz to a shallow NV center and recorded the magnetic spectrum with dynamical decoupling and Qdyne techniques. For dynamical decoupling, an XY8-1 sequence was used where the spacing between π-pulses applied to the NV center was adjusted sequentially, repeated more than $10^6$ times, and the average NV fluorescence at each frequency point recorded. Qdyne was implemented by performing XY8-1 measurements with a fixed delay of 500 ns between π-pulses, and the total spin readout and re-initialisation time adjusted so that $T_L = 9\mu s$. Instead of plotting the average NV fluorescence, the Qdyne spectrum is obtained by storing every readout result in parallel and extracting frequency components by Fourier analysis. A total measurement time of 5000 seconds was used in both measurements. In Fig 3C, the XY8-1 spectrum (upper panel) and Qdyne spectrum (lower panel) are plotted, demonstrating the difference in resolution between the techniques. The resolution of the dynamical decoupling spectrum (300 kHz) is determined by the interaction period $T_s$ whereas the resolution of Qdyne is independent of this timescale, reaching a resolution of 607 µHz which is ultimately determined by the stability of the local oscillator. Although the interaction period is slightly shorter than the coherence time of this NV center, the Qdyne linewidth is 8 orders of magnitude narrower than achievable using dynamical decoupling. We further emphasise that Qdyne does not require any feedback or phase-locking to synchronise the detection sequence to the signal as the initial signal phase is arbitrary.

In Fig. 4 we investigate and compare in detail the signal to noise ratio (SNR), spectral linewidth, and frequency precision of Qdyne and dynamical decoupling magnetometry as a function of measurement time. Spectra obtained with each technique for various measurement times are plotted in Fig. 4A and offset for clarity. Both spectra show raw data without smoothing, but for the dynamically decoupled (XY8-1) spectra Lorentzian fits are included. While the sensitivity of both techniques scales as $T^{1/2}$, i.e. photon shot-noise limited sensitivity, it is apparent that the SNR of Qdyne is superior to conventional XY8 spectroscopy, with a 21 fold improvement in sensitivity (Fig. 4B). The reason for this is twofold. Firstly, Qdyne is more sensitive to weak magnetic fields by virtue of allowing projective readout around an arbitrary axis in order to maximise readout contrast, whereas this is not the case for XY8 spectroscopy (see Fig. 2, supplementary information). As the magnetic field strength reduces, the sensitivity enhancement of Qdyne becomes even more pronounced. Secondly, the SNR of dynamical decoupling spectroscopy decreases as the number of points in the spectrum increases, since less time is spent acquiring data at each frequency point. For the 50 point XY8-1 spectrum shown here, this accounts for a reduction in sensitivity by a factor of 7. Inefficient use of resources is overcome with Qdyne by parallel recording of all frequencies in a given bandwidth using Fourier transform spectroscopy, resulting in a 441-fold speed-up in measurement time.

In Fig. 4C we illustrate in more detail how the resolution gains of Qdyne are achieved. For conventional XY8 spectroscopy, the resolution is given by $1/T_s \geq 1/T_2$, thereby motivating work using the NV nuclear spin as a memory to achieve few Hz resolution spectroscopy given by $1/T_M$ (*11-13*). While the resolution of both dynamical decoupling and quantum memory techniques is fixed during the measurement, the resolution of Qdyne improves linearly with measurement time, eventually saturating due to the stability of the clock which is used. Consequently, after 1 second measurement time a 1 Hz resolution is achieved, even when temporal overheads of sensor readout and preparation are taken into account. At longer times, the resolution is slightly greater than $1/T$, e.g. 2.7 mHz after 500 seconds, which is likely due to Allan deviation of the local oscillator.

In addition to reaching a frequency resolution orders of magnitude beyond that given by the sensor coherence time, the temporal scaling of the Qdyne linewidth has significant impact upon the accuracy of frequency estimation. In fact, while as expected, XY8 spectroscopy scales close to $T^{-1/2}$, the precision obtained with Qdyne scales as $T^{-3/2}$, as long as the local oscillator remains stable (Fig. 4D). A frequency precision beyond the associated linewidth is obtained by performing a least-squares fit of the spectra and plotting the estimated value (with 95% confidence intervals of fit accuracy) as a function of measurement time. Through a theoretical analysis of the frequency precision of each technique we detail how this scaling is achieved. We follow the convention used in amplitude estimation by performing the analysis for the case of a perfect, delta-frequency signal.

Two main ingredients provide the improved performance of Qdyne over dynamic decoupling and quantum memories. Firstly, storing information from each readout onto a hard-drive with practically infinite lifetime overcomes the fragility of quantum sensors and allows the signal to be reconstructed over arbitrarily long periods so a robust clock can be used to its fullest potential. Secondly, Qdyne removes temporal overhead since the sensor continually samples the external field rather than storing phase information on a quantum memory and waiting before continuing with next measurement. From a practical perspective, Qdyne also bypasses the physical overhead of hybrid quantum sensors by removing requirements for additional quantum bits and quantum gates. The precision of frequency estimation, as for amplitude estimation, depends linearly upon signal to noise, but with an additional factor due to the signal linewidth. The first (resolution) ingredient scales linearly with time whereas the second (signal to noise) ingredient, scales as the square root of time. Combining both, a novel precision scaling of $T^{-3/2}$ is achieved.

The precision obtained with dynamical decoupling is given by: $\Delta \nu^{DD} \approx \frac{1}{kT_2\sqrt{TT_2}}$, where the relation can be made exact by including readout parameters. A similar result is obtained when using a memory qubit: $\Delta \nu^M \approx \frac{1}{kT_2\sqrt{TT_M}}$, where $T_M$ is the coherence time of the memory qubit (see supplementary information). Here we have assumed a best case scenario where each point of the spectrum samples the signal and thereby gives an equal contribution to the precision, however this is often not the case, and results in a poorer precision.

For Qdyne, the combination of linewidth and sensitivity gives a precision: $\Delta \nu^Q = \frac{1}{kT\sqrt{TT_2}}$, valid for measurement times up to the stability of the local oscillator. The factors $T, \sqrt{T}$ and $\sqrt{T_2}$ originate respectively from the regular inverse time scaling of resolution, (which in previous methods was limited by the coherence time of the quantum probes (*16-21*)), the signal to noise which improves as the square root of the number of measurements, and a rotation of the sensor by a larger angle as the interaction is prolonged. For measurements times exceeding the clock stability $T_{LO}$, the precision of Qdyne follows the standard quantum limit $\Delta \nu^Q = \frac{1}{kT_{LO}\sqrt{TT_2}}$ (Fig. 4D).

A comparison between the ultimate precision limits of Qdyne and dynamic decoupling/quantum memory techniques gives: $\frac{\Delta \nu^{DD}}{\Delta \nu^Q} \approx \frac{T_{LO}}{T_2}$, and $\frac{\Delta \nu^M}{\Delta \nu^Q} \approx \frac{T_{LO}}{\sqrt{T_2 T_M}}$, where all relations have been derived in the limit of a weak signal $kT_2 \ll 1$, i.e. the sensor rotates around a small angle. Under this condition, and given $T_{LO} \gg T_2, T_M$, significant improvement in precision can be achieved with Qdyne; as much as 9 orders of magnitude in comparison to dynamic decoupling (Fig. 4D). The final absolute precision of $2\times10^{-5}$ Hz reached with Qdyne here, corresponds to a frequency estimation of the 1 MHz signal at the eleventh fractional digit ($\Delta \nu/\nu \approx 10^{-11}$).

Due to the rigorous nature of the Heisenberg limit in quantum measurement, claims of a precision scaling beyond either the $T^{-1}$ or $n^{-1}$ warrant further discussion. Indeed one may take the stance that Qdyne continuously samples an oscillating field, therefore the entire sequence should be treated as a single measurement. From this classical signal analysis perspective, the Cramer-Rao bound yields the same precision scaling as $\Delta \nu^Q$ (*27*). From the alternate quantum metrology framework, where each readout of the quantum sensor constitutes a measurement, then a $T^{-3/2}$ (or $n^{-3/2}$) scaling in terms available resources is achieved. Of further interest is that such a precision scaling is achieved when each measurement duration is fixed at $T_2$, i.e. the absolute precision is maximised, so that no leveraging of dynamical range occurs as is the case for adaptive feedback schemes (*28*). Further analysis of the Fisher Information indeed gives a super-linear precision scaling and in the case of infinitely long probe coherence time, a Heisenberg limited precision of $T^{-2}$ is obtained (see supplementary information and Ref. (*29*)). As such, Qdyne does not break the Heisenberg limit, but takes advantage of the extra parameter space afforded to frequency measurements in quantum metrology.

In Fig. 4E we highlight the FFT spectroscopy mode of Qdyne, where all spectral components in a given bandwidth are acquired simultaneously. In the context of acquiring sparse spectra, FFT spectrometers offer significant enhancement over 'frequency-swept' spectrometers since the time spent collecting no data (dead-time) is removed. To demonstrate, we recorded the spectrum from three independent signal generators with 7 mHz resolution, over a measurement duration of 180 seconds. For the measured frequency differences of 22 Hz and 25 Hz, approximately 10,000 datapoints span the spectrum. Obtaining the same signal to noise using conventional methods, is prohibitively long taking an estimated 500 hour duration (see ref (*12*) for comparison).

The bandwidth of the Qydne technique can be adjusted by tuning $T_L$ as it determines the sampling rate, which sets an upper limit on the detection frequency. For the case that the sensor readout time is a small fraction of the signal accumulation time, the detection bandwidth closely approximates the filter function of the implemented XY8 sequence. In Fig. 4F we plot a series of Qdyne spectra recorded with an XY8-1 sequence centerd at 1 MHz ($\tau = 500$ ns), where the signal frequency was recorded for 100 seconds for a series of frequency detunings. The recorded signal amplitude follows the expected behaviour given by the XY8 filter function (plotted in green), until a detuning of several kHz. The earlier than expected decay is currently not well understood, but appears to be related to a relative timing instability of the external clock leading to a broadening of the signal. Linewidth broadening reduces the signal amplitude but not the total power, by integrating the area of each curve (purple points), the expected bandwidth of the Qdyne technique is recovered. There is an inherent trade-off between magnetic sensitivity and bandwidth, since magnetic sensitivity can be increased by extending each interaction period $T_s$ (up to a maximum value of $T_2$), but this comes with a commensurate reduction in bandwidth. Likewise the sensitivity can be increased by spending more time reading out the sensor, with a decrease in bandwidth due to a lower sampling rate.

Whilst measurements of a narrow frequency standards demonstrate the resolution enhancement of Qdyne, until this point the magnetic sensitivity of the quantum sensor has not been fully utilised. We now apply Qdyne to nanoscale nuclear magnetic resonance spectroscopy, based on magnetic dipole interaction between the NV electronic spin and statistically polarized molecules near the diamond surface (Fig. 5A). Due to the short range nature of the dipolar interaction, the majority of the signal arises from a volume of $\sim 2d^3$, where $d$ is the NV depth to the surface. For the 5 nm deep NV used here, this corresponds to a detection of $\sim 4 \times 10^4$ protons, which is below the sensitivity limit of conventional NMR spectrometers.

Using Qdyne we measured the proton NMR spectrum from a sample of polybutene placed on the diamond surface, for 7 hours. The recorded photon time trace was split into time intervals which were tuned to optimise the SNR dependant on the correlation time of the signal. For the estimated $10^{-15}$ m$^2$/s diffusion coefficient of polybutene, we can determine how the spin-spin relaxation time and interaction time with the NV center $\tau_C^{NV}$ (inversely proportional to the diffusion coefficient) affect the signal correlation time. In particular, we expect the correlation time to be determined primarily by proton spin-spin relaxation in the viscous oil of timescales between $0.1 - 0.01$ ms. The measured linewidth of 2 kHz, agrees with this analysis. We stress here that the local oscillator still defines measurement resolution, and the recorded linewidth is due to the physical parameters of the sample. The recorded spectrum is well replicated by performing a rigorous atomistic simulation of a 6x12x12 nm$^3$ volume, corresponding to 70304 diffusing nuclear spins interacting with the NV center, and neglecting internuclear correlations (see supplementary information). Using the same experimental sensing time and diffusion coefficient, with a proton relaxation time of 0.32 ms the physical box provides converged data as the interaction with more distant particles is effectively suppressed by the finite T$_1$ time.

We have introduced a new spectroscopic technique which references a sensitive qubit to a stable oscillator, allowing a resolution beyond the $T_2$ limit of the quantum sensor to be reached. The technique surpasses the standard quantum limit for identical and independent measurements to achieve a $T^{-3/2}$ frequency precision scaling, which has not previously been demonstrated in quantum estimation scenarios. Qdyne also removes temporal overhead by sampling multiple frequencies in parallel, so that both absolute sensitivity and frequency precision can be increased by orders of magnitude over step-wise frequency sampling. By applying Qdyne to shallow NV centers in diamond we achieved an increase in frequency precision and resolution by 9 and 8 orders of magnitude respectively over dynamic decoupling. The application of Qdyne to nanoscale NMR spectroscopy, where gains in frequency precision translate to spatial resolution has implications for molecular structural determination. Finally, the low technical demands of Qdyne, which bypasses the need for hybrid quantum systems make the technique widely applicable for high resolution magnetic spectroscopy.

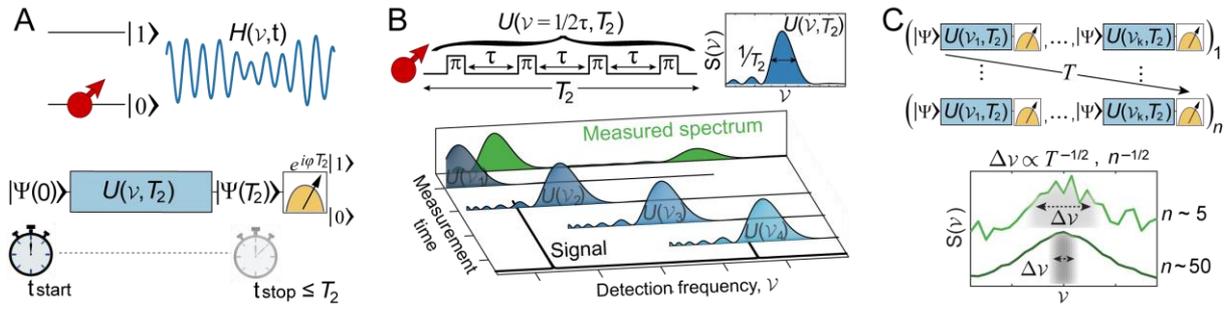

**Fig. 1**. Frequency estimation using a quantum sensor. (**A**) Quantum metrology with a two level system. A spin qubit is used to measure frequency components of a time varying Hamiltonian. The interaction period is limited by the qubit coherence time, after which the qubit state is readout. (**B**) Conventional quantum spectroscopy. A spectral filter is generated during the interaction period by pulsing or driving the sensing qubit. For the XY8 sequence, the filter is centered near frequency $1/2\tau$, where $\tau$ is the interpulse delay, and has a width inversely proportional to the interaction time. The filter is scanned stepwise during multiple measurements in order to acquire a complete spectrum. (**C**) A single interaction period at each detection frequency is repeated multiple times to provide a better frequency estimate. The uncertainty in frequency precision usually scales as square-root the number of measurements.

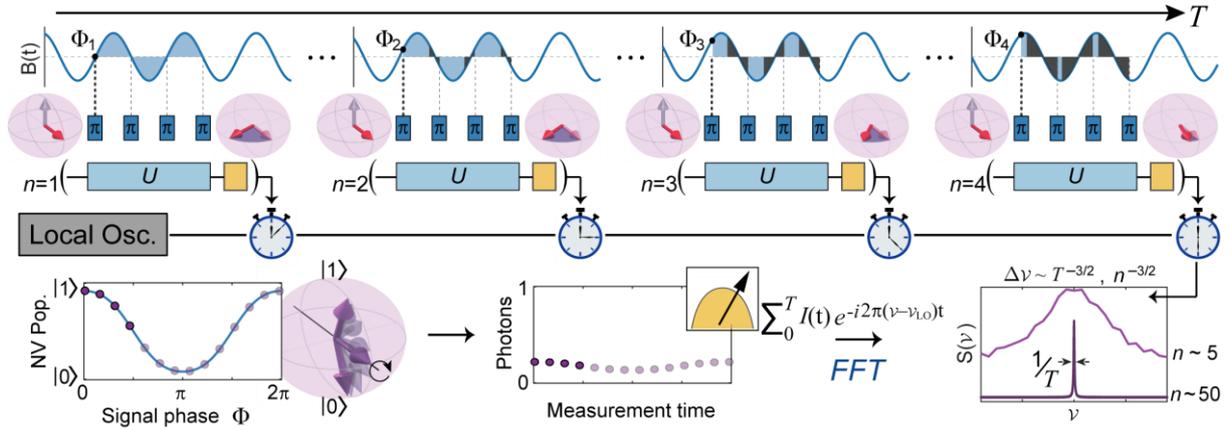

**Fig. 2**. Qdyne detection scheme. The output state of the sensor qubit depends on initial phase $\Phi_1$ of the signal field. The measurement result is recorded and synchronised with an external clock before next measurement is performed for a second signal phase $\Phi_2$. This procedure is repeated for the entire measurement time $T$. By heterodyning with an external clock, the NV population and therefore measured photon intensity records the signal phase evolution in time. A Fourier transform then allows the signal frequency, with respect to local oscillator frequency to be determined. For Qdyne detection, the precision of frequency estimation scales as $T^{-3/2}$.

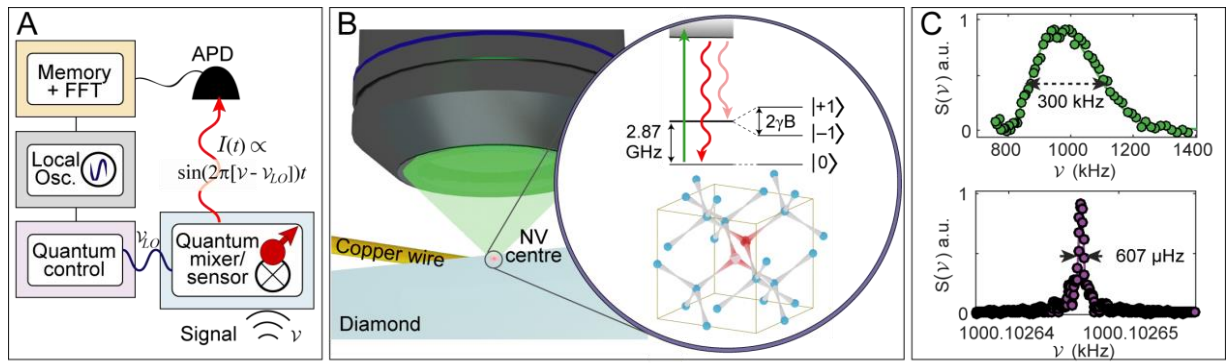

**Fig. 3**. Experimental setup and high resolution spectroscopy of magnetic fields with quantum clock-in detection. (**A**) Schematic of Qdyne experiment. The clock from a local oscillator is mixed with the detected frequency by the quantum sensor (NV spin) and the intensity modulated fluorescence recorded. (**B**) The experimental setup. Confocal microscopy of single NV centers (~5nm from the diamond surface) is performed with 532nm excitation. A copper wire is used for microwave and radio frequency control. (Inset) Green excitation of the NV center optically pumps the NV spin into the $|0\rangle$ spin state. The $|0\rangle$ state fluorescence stronger than the $|\pm 1\rangle$ states, allowing the NV spin to be read-out optically. The physical structure of the NV center is a nitrogen atom adjacent to a vacancy in the diamond lattice. (**C**) Spectroscopy of magnetic fields with a single NV center. (Upper panel) Conventional XY8 magnetometry of a 1 MHz magnetic field resulting in a linewidth of 300 kHz. (Lower panel) Qdyne detection of the same magnetic field yields a spectrum with a linewidth of 607 µHz, limited by the stability of a quartz crystal oscillator connected to the experiment.

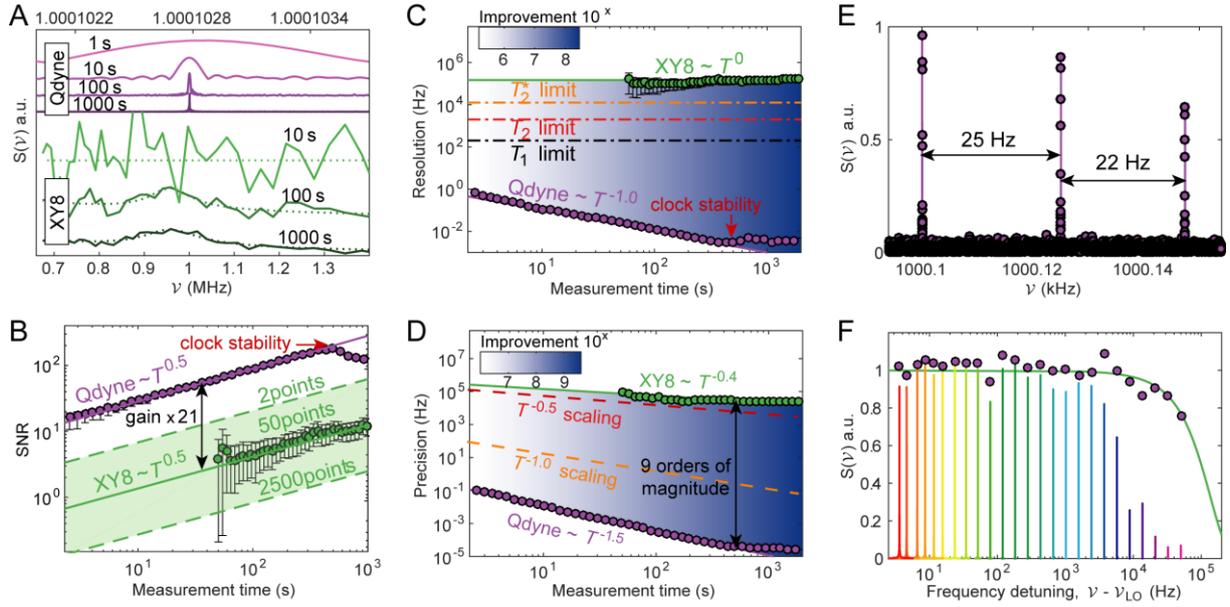

**Fig. 4**. Performance of Qdyne (and conventional XY8) spectroscopy. (**A**) Magnetic spectra recorded with Qdyne (upper panel, 1 – 1000s measurement times) and XY8 (lower panel, 10 – 1000s measurement times) techniques. Lorentzian fits to XY8 data are shown as dotted lines. (**B**) Temporal scaling of signal-to-noise ratio (SNR) and magnetic field sensitivity of Qdyne and XY8 magnetometry. Both techniques show photon shot-noise limited sensitivity, with SNR increasing as $\sqrt{T}$. At measurement times longer than the external clock stability of ~500s, the SNR of Qdyne is reduced. (**C**) Spectral resolution of Qdyne and XY8 techniques versus measurement time. The full-width-half-maximum (FWHM) of the signal measured with XY8 magnetometry is set by the interaction time, and remains constant at 300 kHz for the measurement duration. The resolution of Qdyne improves linearly with measurement time, including overheads of sensor readout and preparation time, until the external clock stability is reached. (**D**) Precision of frequency estimation for Qdyne and XY8 techniques as a function of measurement time, determined by least-squares fitting of the signal peak (see supplementary information). While the precision of XY8 magnetometry improves near the standard quantum limit, i.e. $T^{-1/2}$ as expected for independent measurements, Qdyne display a precision improving as $T^{-3/2}$, see text for discussion. (**E**) Fast Fourier transform spectroscopy of multiple fields using Qdyne. Simultaneous detection of magnetic fields at three different frequencies (separated by 25 Hz and 22 Hz), and generated by three independent signal generators. (**F**) Frequency bandwidth of Qdyne. The sensitivity to magnetic fields of different detuning is plotted in comparison to the XY8-1 filter function (green line). Each spectrum, shown with a different color, is recorded for the same amount of time. The total integrated signal for each detuning is plotted with a purple circle.

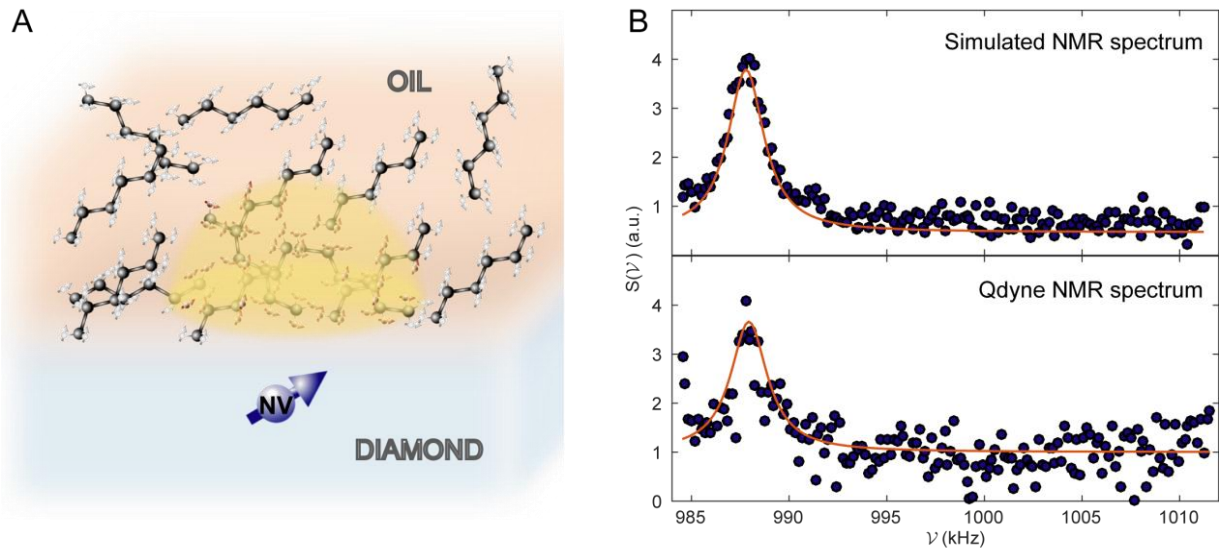

**Fig. 5**. Nanoscale NMR spectroscopy of polybutene with an NV center in diamond. (**A**) Sketch of the Qdyne NMR experiment. The NV electronic spin interacts with proton spins in polybutene molecules placed on diamond surface. The correlation time of the signal is determined by diffusion of molecules through the detection volume (yellow), and dipole-dipole interaction between protons in the sample. (**B**) Qdyne detection of proton Larmor frequency. Measured and simulated proton spectrum of polybutene using Qdyne with a 5 nm deep NV center. Solid lines are Lorentzian fits.